# GPGPU PROCESSING IN CUDA ARCHITECTURE

Jayshree Ghorpade[1], Jitendra Parande[2], Madhura Kulkarni[3], Amit Bawaskar[4]

[1]Departmentof Computer Engineering, MITCOE, Pune University, India
jayshree.aj@gmail.com
[2] SunGard Global Technologies, India
jitendra.parande@sungard.com
[3] Department of Computer Engineering, MITCOE, Pune University, India
madhurak25@gmail.com
[4]Departmentof Computer Engineering, MITCOE, Pune University, India
amitbawaskar@gmail.com

## ABSTRACT

*The future of computation is the Graphical Processing Unit, i.e. the GPU. The promise that the graphics cards have shown in the field of image processing and accelerated rendering of 3D scenes, and the computational capability that these GPUs possess, they are developing into great parallel computing units. It is quite simple to program a graphics processor to perform general parallel tasks. But after understanding the various architectural aspects of the graphics processor, it can be used to perform other taxing tasks as well. In this paper, we will show how CUDA can fully utilize the tremendous power of these GPUs. CUDA is NVIDIA's parallel computing architecture. It enables dramatic increases in computing performance, by harnessing the power of the GPU. This paper talks about CUDA and its architecture. It takes us through a comparison of CUDA C/C++ with other parallel programming languages like OpenCL and DirectCompute. The paper also lists out the common myths about CUDA and how the future seems to be promising for CUDA.*

## 1. INTRODUCTION

GPU computation has provided a huge edge over the CPU with respect to computation speed. Hence it is one of the most interesting areas of research in the field of modern industrial research and development.

GPU is a graphical processing unit which enables you to run high definitions graphics on your PC, which are the demand of modern computing. Like the CPU (Central Processing Unit), it is a single-chip processor. However, as shown in Fig. 1, the GPU has hundreds of cores as compared to the 4 or 8 in the latest CPUs. The primary job of the GPU is to compute 3D functions. Because these types of calculations are very heavy on the CPU, the GPU can help the computer run more efficiently. Though, GPU came into existence for graphical purpose, it has now evolved into computing, precision and performance.

The evolution of GPU over the years has been towards a better floating point performance. NVIDIA introduced its massively parallel architecture called "CUDA" in 2006-2007 and changed the whole outlook of GPGPU programming. The CUDA architecture has a number of processor cores that work together to munch the data set given in the application. GPU computing or GPGPU is the use of a GPU (graphics processing unit) to do general purpose scientific and engineering computing. The model for GPU computing is to use a CPU and GPU together in a heterogeneous co-processing computing model. The sequential part of the application runs on the CPU and the computationally-intensive part is accelerated by the GPU. From the user's point of view, the application is faster because it is using the better performance of the GPU to improve its own performance.

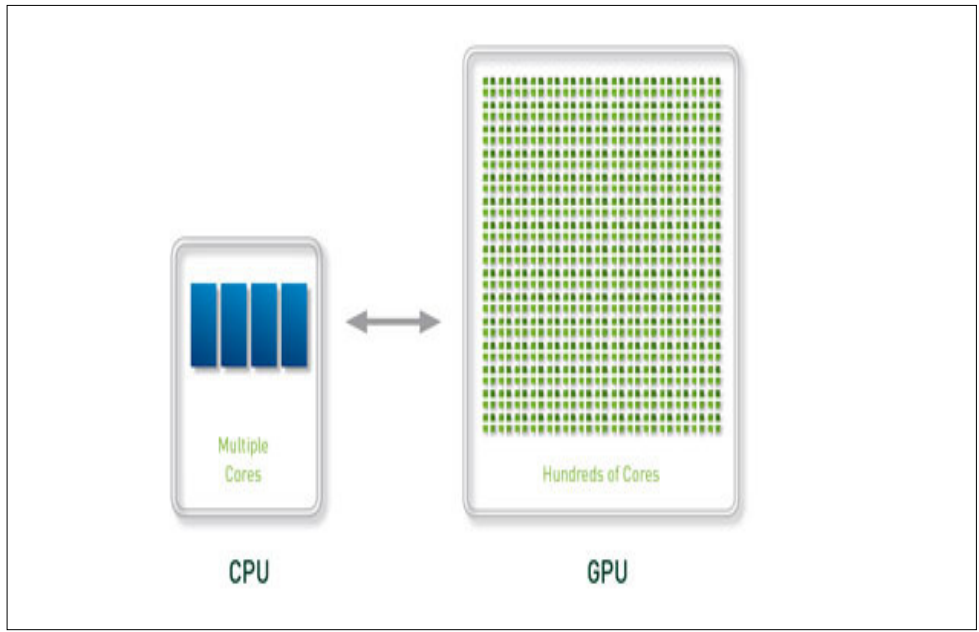

Fig.1: Core comparison between CPU and GPU [1]

All programming instructions help us to derive necessary data, done with the help of CPU. Now a days CPU is capable of crunching more numbers than before. Still processing huge data or crunching large numbers puts a lot of burden on CPU. This can be done powerfully using Graphics Processing Unit. If we do complex calculations using GPU then we can free CPU time cycles which can be used for other higher priority tasks. In such a way GPU can assuage the burden on CPU by handling all the intense mathematical calculations.

Earlier, CPUs used to handle all of the computations and instructions in the whole computer. But as technology progressed, it became advantageous to take out some of the tasks from the CPU and have it performed by other microprocessors. As CPU processes data centrally it is known as Central Processing Unit. But as technology progressed, CPU is now able to transform some of its duties to be done with the help of other microprocessors.

The GPU is a device that is beneficial primarily to people that has intensive graphical functions on their computer. In other words, if you just use Microsoft Office and the e-mail page of your browser when you are on the computer, chances are very good that the GPU will not add that much to your computing experience. However, if you play video games and look at videos on the internet on a frequent basis, what you will discover is that installing a GPU onto your computer will greatly improve the performance you get out of the entire thing. Improving computer performance is always a good thing and this is why the GPU has become very popular in recent years. GPU computing is on the rise and continuing to grow in popularity and that makes the future very friendly for it indeed.

The comparison of the CPU and GPU is shown in table 1. As we can see, the CPU is more efficient for handling different tasks of the Operating systems such as job scheduling and memory management while the GPU's forte is the floating point operations.

| CPU | GPU |
|---|---|
| –Really fast caches (great for data reuse) | – Lots of math units |
| – Fine branching granularity | – Fast access to onboard memory |
| – Lots of different processes/threads | – Run a program on each fragment/vertex |
| – High performance on a single thread of execution | – High throughput on parallel tasks |
| CPUs are great for task parallelism | GPUs are great for data parallelism |
| CPU optimised for high performance on sequential codes (caches and branch prediction) | GPU optimised for higher arithmetic intensity for parallel nature (Floating point operations) |

Table 1: Comparison between CPU and GPU

## 2. GPGPU

GPU functionality has, traditionally, been very limited. In fact, for many years the GPU was only used to accelerate certain parts of the graphics pipeline. The processing capacity of the GPU is limited to independent vertices and fragments. But this processing can be done in parallel using the multiple cores available to the GPU. This is especially effective when the programmer wants to process many vertices or fragments in the same way. In this sense, GPUs are processors that can operate in parallel by running a single kernel on many records in a stream at once. A stream is simply a set of records that require similar computation. Streams provide data parallelism. We can multiple inputs and outputs at the same time, but we cannot have type of memory which is not only readable but also writeable. This enables the GPU to have Data-parallel processing, as the GPU architectures are ALU-heavy and contain multiple vertex & fragment pipelines. This results in tremendous computational power for the GPU. But this computational capability was unutilised in the earlier evolutionary stages of the GPU.

Using the GPU for processing non graphical entities is known as the General Purpose GPU or GPGPU.Traditionally GPU was used to provide better graphical solutions for available environments. If we use GPU for computationally intensive tasks, then this kind of work is known as GPGPU. It is used for performing complex mathematical operations in parallel for achieving low time complexity. The arithmetic power of the GPGPU is a result of its highly specialised architecture. This specialisation results in intense parallelism which can lead to great advantages if used properly. But this architecture comes at a price. As a result of the parallelism the GPGPU cannot handle integer data operands, bit shift and bitwise operations, double precision arithmetic. It is an unusual model for a processor. Such difficulties are intrinsic nature of the graphics card hardware and not a result of immature technology. Use of GPU for the tasks other than graphical purpose is made possible by additional programmable stages. It also specializes in floating point operations. Higher precision arithmetic also made possible use of GPU for non-graphical data.

As a result of the irresistible qualities of the GPU an increasing number of programmers are trying to make their application more and more GPGPU oriented. Advantages of GPU programmability are so irresistible that many programmers have begun rewrite their applications using GPGPU. In that way they can make use of high computational capacity of GPU. The biggest hurdle that a programmer faces when first programming a GPGPU is learning how to get the most out of a data-parallel computing environment. GPU parallelism is present at more than one level. Parallel execution on multiple data elements is a fundamental design element of GPGPUs.

Parallelism is obtained be rearranging the data from scalar quantities to vector ones. The vector processors can schedule more than one instruction per cycle. And so, the vectors can be computed in parallel if their elements are independent of each other. This results in a chance for instruction level parallelism, which the GPU can utilize.

There are multiple SDKs and APIs available for the programming of GPUs for general purpose computation that is basically other than graphical purpose for example NVIDIA CUDA, ATI Stream SDK, OpenCL, Rapidmind, HMPP, and PGI Accelerator. Selection of the right approach for accelerating a program depends on a number of factors, including which language is currently being using, portability, supported software functionality, and other considerations depending on the project.

At present CUDA is the predominant method for GPGPU acceleration, although it is only supported by NVIDIA GPUs. In the longer term OpenCL promises to become the vendor-neutral standard for programming heterogeneous multi-core architectures. As it is largely based on CUDA, there will hopefully be relatively little difficulty making the switch from CUDA's proprietary API to OpenCL.

## 3. CUDA

CUDA (Compute Unified Device Architecture) is NVIDIA's GPU architecture featured in the GPU cards, positioning itself as a new means for general purpose computing with GPUs [3]. CUDA C/C++ is an extension of the C/C++ programming languages for general purpose computation. CUDA gives advantage of massive computational power to the programmer. This massive parallel computational power is provided by Nvidia's graphics cards.

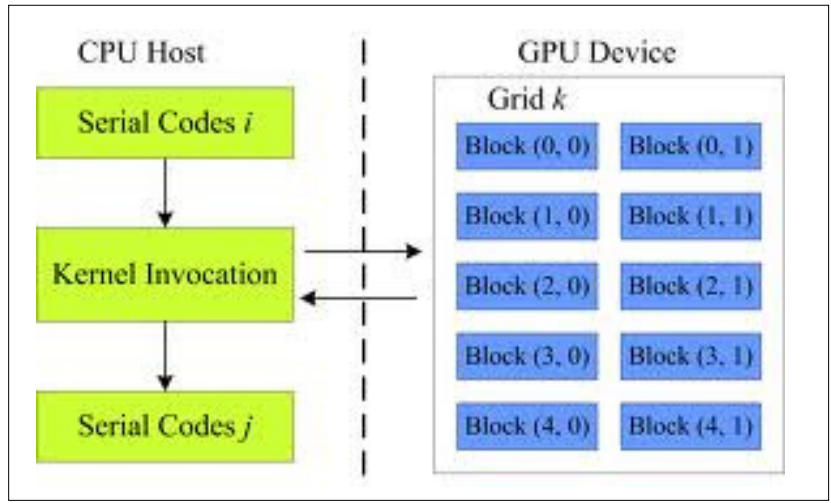

Fig.2: Flow of execution of GPU [2]

CUDA provides 128 co-operating cores. For running multithreaded applications there is no need of streaming computing in GPU, because cores can communicate also can exchange information with each other. CUDA is only well suited for highly parallel algorithms and is useful for highly parallel algorithms. If you want to increase performance of your algorithm while running on GPU then you need to have many threads. Normally more number of threads gives better performance. For the most of the serial algorithms, CUDA is not that useful. If the problem cannot be broken down into at least a thousand threads then using CUDA has no overall advantage. In that case we can convert serial algorithm into parallel one but, it is not always possible. As mentioned above to get best optimization you need to divide your problem into minimum thousand threads. Then performance of algorithm increases rapidly.

CUDA can be taken full advantage of when writing in C. As stated previously, the main idea of CUDA is to have thousands of threads executing in parallel. All of these threads are going to be executing the very same function (code), known as a kernel. All these threads are executed using the same instruction and different data. Each thread will know its own ID, and based off its ID, it will determine which pieces of data to work on.

A CUDA program consists of one or more phases that are executed on either the host (CPU) or a device such as a GPU. As shown in Fig. 2 in host code no data parallelism phase is carried out. In some cases little data parallelism is carried out in host code. In device code phases which has high amount of data parallelism are carried out. A CUDA program is a unified source code encompassing both, host and device code. The host code is straight forward C code. In next step it is compiled with the help of standard C compiler only. That is what we can say ordinary CPU process. The device code is written using CUDA keywords for labelling data-parallel functions, called kernels, and their associated data structures. In some cases one can also execute kernels on CPU if there is no GPU device available. This facility is provided with the help of emulations features. CUDA software development kit provides these features.

One advantage is that there is no need to write the whole programme using CUDA technology. If you are writing a large application, complete with a user interface, and many other functions, and then most of your code will be written in C++ or whatever your language of choice is. When you really want to do large mathematical computations then, you can simply write kernel call to call CUDA functions you have written. In this way instead of writing complete programme you can use GPU for some portion of the code where we need huge mathematical computations.

## 4. ARCHITECTURE

GPU is a massively parallel architecture. Many problems are such that they can be efficiently solved using GPU computing. GPUs have large amount of arithmetic capability. They increase the amount of programmability in the pipeline.

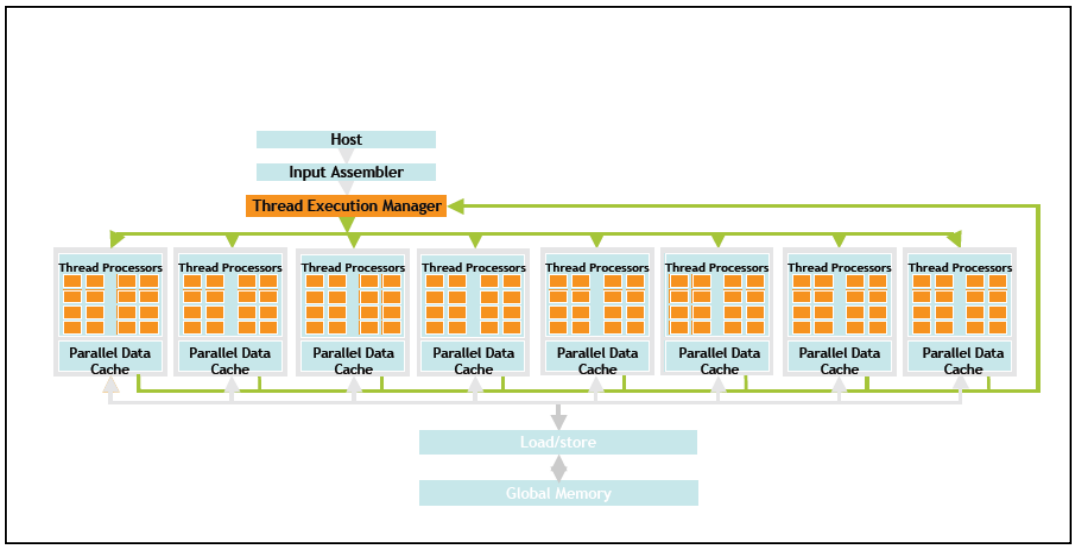

Fig.3: GPU Architecture [4]

Fig. 3 shows the architecture of a typical CUDA-capable GPU. CUDA can be seen to be an array of streaming processors capable of high degree of threading. In Fig. 3, two SMs form a building block; however, the number of SMs in a building block can vary from one generation of CUDA GPUs to another generation. Also, each SM in Fig. 4 has a number of streaming processors (SPs) that share control logic and instruction cache. Each GPU currently comes with up to 4 gigabytes of graphics double data rate (GDDR) DRAM, referred to as global memory in Fig. 3. These RAMs of the GPU are different than the CPU as they are used as frame buffer memories used for rendering graphics. For graphics applications, they hold video images, and texture information for three-dimensional (3D) rendering, but for computing they function as very-high-bandwidth, off-chip memory, though with somewhat more latency than typical system memory. For massively parallel applications, the higher bandwidth makes up for the longer latency.

### 4.1 Basic Units of CUDA [4]

CUDA Architecture comprises of three basic parts, which help the programmer to effectively utilise the full computational capability of the graphics card on the system. CUDA

Architecture splits the device into grids, blocks and threads in a hierarchical structure as shown in fig. 4. Since there are a number of threads in one block and a number of blocks in one grid and a number of grids in one GPU, the parallelism that is achieved using such a hierarchical architecture is immense.

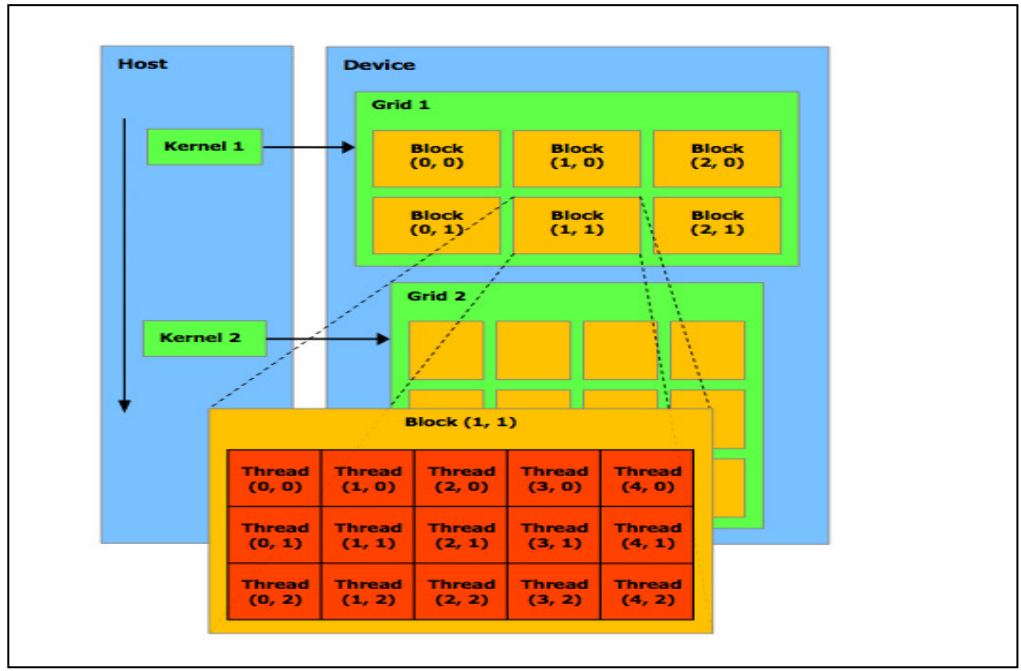

Fig. 4: CUDA Architecture [4]

### 4.1.1 The Grid

A grid is a group of threads all running the same kernel. These threads are not synchronized. Every call to CUDA from CPU is made through one grid. Starting a grid on CPU is a synchronous operation but multiple grids can run at once. On multi-GPU systems, grids cannot be shared between GPUs because they use several grids for maximum efficiency.

### 4.1.2 The Block

Grids are composed of blocks. Each block is a logical unit containing a number of coordinating threads, a certain amount of shared memory. Just as grids are not shared between GPUs, blocks are not shared between multiprocessors. All blocks in a grid use the same program. A built in variable "blockIdx" can be used to identify the current block. Block IDs can be 1D or 2D (based on grid dimension). Usually there are 65,535 blocks in a GPU.

### 4.1.3 The Thread

Blocks are composed of threads. Threads are run on the individual cores of the multiprocessors, but unlike grids and blocks, they are not restricted to a single core. Like blocks, each thread has an ID (threadIdx). Thread IDs can be 1D, 2D or 3D (based on block dimension). The thread id is relative to the block it is in. Threads have a certain amount of register memory. Usually there can be 512 threads per block.

## 4.2 CUDA Memory types:-

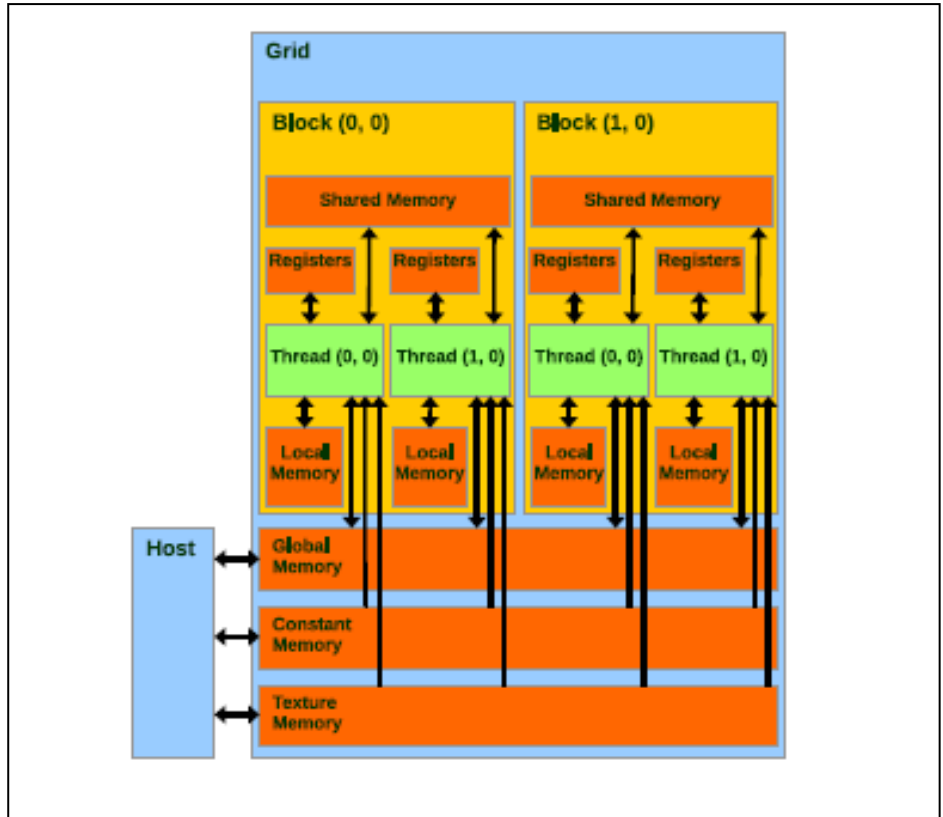

Fig.5: CUDA structure [5]

### 4.2.1 Global memory:-

It is a read and write memory. It is slow & uncached and requires sequential & aligned 16 byte reads and writes to be fast (coalesced read/write).

### 4.2.2 Texture memory:-

It is a read only memory. Its cache optimized for 2D spatial access pattern.

### 4.2.3 Constant memory:-

This is where constants and kernel arguments are stored. It is slow, but with cache.

### 4.2.4 Shared memory:-

All threads in a block can use shared memory for read or write operations. It is common for all threads in a block and its size is smaller than global memory. The number of threads that can be executed simultaneously in a block is determined the shared memory that is specified and it denotes the occupancy of that block.

### 4.2.5 Local memory:-

It is generally used for whatever does not fit into registers. It is slow & uncached, but allows automatic coalesced reads and writes.

**4.2.6 Registers:-**

This is likely the fastest memory available. One set of register memory I given to each thread and it uses them for fast storage and retrieval of data like counters, which are frequently used by a thread.

## 5. RELATED WORK

Many researchers have realized the power and capabilities of the GPU. So a lot of work is being done now days to implement GPU into existing technologies and algorithms, to enhance their performance. Parallel computing has not only helped increase speed and resource utilization but has also made many difficult computations possible. Due to all these advantages it has been named among "The Top Trends in High Performance Computing" [12]

Of all these amazing advances in the field we are stating a few.

- The recent developments in the field of GPUs have brought a relevant amount of computing power to computers which provides a way to accelerate several numerical electromagnetic methods. Danilo De Donno et al [13] explain how to exploit GPU features by examining how the computational time of the Finite-Difference Time-Domain (FDTD) Method can be reduced. Numerical results obtained from two-dimensional study of a human-antenna interaction problem are used to demonstrate efficiency. Accordingly, they explain how to implement a CUDA version of the FDTD algorithm so as to achieve peak performance. [13]

- To implement parallel programming on GPUs using C, CUDA is one of the simple ways. David Luebke [14] examines this approach to perform matrix addition on GPUs using CUDA C. Dividing the different loop variables into threads, grids and blocks results into effective implementation. He also gives a view of the different memories present in the GPU and how to use them in programming. In conclusion, he tries to give a feel of what the GPU can do on a small introductory level of implementation and how this implementation helps to speed up a calculation as compared to a CPU.

- Miguel Cárdenas-Montes et al [15] try to solve large size optimization problems using GPUs. They examine the shortcomings existing in the current optimization problem solving methods including limited time budget among other factors. They describe the design of a GPGPU-based Parallel Particle Swarm Algorithm, to tackle this type of problem maintaining a limited execution time budget. This implementation profits from an efficient mapping of the data elements (swarm of very high dimensional particles) to the parallel processing elements of the GPU. In this problem, the fitness evaluation is the most CPU-costly routine, and therefore the main candidate to be implemented on GPU. As main conclusion, the speed-up curve versus the increase in dimensionality is shown.

- The GPU is being used to help the actuaries crunch their data faster to obtain good insurance policies. The process of calculating the rate of insurance fluctuates according to many parameters for different policy holders. These parameters range from a simple study of maturity period of the insurance to a very complex data collection of the mortality rate. As more and more parameters are added the calculation becomes more and more complex. The final result obtained by computation on CPU may not be cost effective using CPU as a lot of resources are wasted and time taken to perform these calculations is huge. This is where GPU computation comes in handy. It can help increase the speed of computation of a simple desktop machine by adding a GPU processor to aid the CPU for these complex calculations. The GPU can work on different parameters in parallel, and as a result the final result can be obtained using a single pass of the GPU.

Thus we see the amount of work that is going on in the field of GPGPU is exemplary and looking at the current trends this is going to increase. The GPGPUs parallel computing capability has opened new doors in the computing world. It has helped desktop PCs overcome the traditional bottleneck of the CPU and as a result increase the overall capacity of the desktop PC. As we see from the above cited examples that the application of GPGPU is not limited to just one field. It can be applied to a variety of domains and can help computers do things which were not possible a few years back.

## 6. EXPERIMENTAL SETUP

The Software Development Kit or the SDK is a good way to learn about CUDA, one can compile the examples and learn how the tool kit works. The SDK is available at NVidia's website and can be downloaded by any aspiring programmers who want to learn about CUDA programming. Anyone who has some basic knowledge on C programming can start CUDA programming very quickly. No prior knowledge of graphics programming is required to write CUDA codes. CUDA is derived from C with some modifications which enabled it to run on GPU. CUDA is C for GPU.

**6.1 To start developing your own applications in CUDA follow the following steps:**

1. Install visual studios as an environment for CUDA programming.

2. Install specific GPU drivers according to Graphics card and also install the CUDA SDK according to the compute capability of your Graphics card.

3. Write the program code according to normal C/C++programming constructs.

4. Now, change the written program into the CUDA parallel code by using the library functions provided by the SDK. The library functions are used to copy data from host to device, change execution from CPU to GPU and vice versa, copy data from device to host.

**6.2 The basics of CUDA code implementation as:**

1. Allocate CPU memory.

2. Allocate same amount of GPU memory using library function "CudaMalloc".

3. Take data input in CPU memory.

4. Copy data into GPU memory using library function CudaMemCpy with parameter as (CudaMemcpyHostToDevice)

5. Perform processing in GPU memory using kernal calls (kernel calls are a way to transfer control from CPU to GPU; they also specify the number of grids, blocks and threads i.e. the parallelism required for your program.

6. Copy final data in CPU memory using library function CudaMemCpy with parameter as (CudaMemcpyHostToDevice)

7. Free the GPU memory or other threads using library function CudaFree.

As we can see, setting up the environment and writing code in CUDA is a fairly easy task. But, it requires that the programmer must have good know-how of the architecture and knowledge of writing parallel codes. The most important phase of programming in CUDA is the kernel calls wherein the programmer must determine the parallelism that the program requires. The division of data into appropriate number of threads is the major area which can make or break a code.

## 7. COMPARATIVE STUDY OF DIFFERENT ARCHITECTURES

### 7.1 Direct Compute

Microsoft DirectX is a collection of Application Programming interfaces (APIs) for handling tasks related to multimedia, especially game programming and video, on Microsoft platforms. [7]

*Pros:-*

1. API is very easy to learn for programmers with some knowledge of DX.

2. Efficient interoperability with D3D graphics resources.

3. Can be easily used in combination with the current games which employ DirectX.

4. Access to texturing features

5. Works with most DX10 and DX11 GPUs

*Cons:-*

1. Windows 7 and Vista support only.

2. Lack of libraries, examples and documentation

3. Fewer bindings available

4. No CPU fallback

### 7.2 OpenCL

OpenCL (Open Computing Language) is a framework for writing programs that execute across heterogeneous platforms consisting of CPUs, GPUs and other processors. It is an open standard derived by Khronos Group. [8]

*Pros:-*

1. Can be implemented on number of platforms and supports wider range of hardware. Supports AMD, Nvidia and Intel GPUs equally. Can also be used on latest versions of cell phones and other electronic devices.

2. When GPU hardware is not present then it can fall back on to CPU to perform the specified work.

3. Supports synchronization over multiple devices.

4. Easy to learn, as one has to just add OpenCL kernel in code.

5. An open standard and not vendor locked.

6. Share resources with OpenGL

*Cons:-*

1. Still public drivers support for OpenCL 1.1 is not present.

2. Lacks mature libraries

3. CUDA has more advance debugging and profiling tools than OpenCL.

## 7.3 CUDA

CUDA stands for Compute Unified Device Architecture. It is a parallel computing architecture developed by Nvidia. CUDA is the computing engine provided by Nvidia. Graphics processing units (GPUs) are accessible to software developers through number of industry standard programming languages other than CUDA such as OpenCL, DirectX. [9]

*Pros:-*

1. The kernel calls in CUDA are written in simple C-like languages. So, programmers task is simpler.

2. Kernel code has full pointer support

3. Supports C++ constructs.

4. Fairly simple integration API.

5. Better fully GPU accelerated libraries currently available.

6. CUDA can be used for a large number of languages.

7. The programmer has a lot of elp available in the form of documentation and sample codes for different platforms.

8. A programmer can use CUDA's visual profiler, a tool used for performance analysis.

9. Updates are more regular.

10. Has been on the market much longer

*Cons:-*

1. Restricted to Nvidia GPUs only

2. If CUDA accelerated hardware is not present, error is reported and task is aborted. There is no CPU backup in this scenario.

3. It is difficult to start programming in CUDA, because it requires setting up of the environment in which CUDA enabled programs can run. For example the NVCC compiler has to be incorporated into build.

The above subsections give us a good idea of the different architectures and programming languages that we can use at present. Considering the factors mentioned below, we can say that CUDA is better because of the following reasons:-

1. It is flexible as it can be used with diverse platforms.

2. Large number of documentation and examples available.

3. Full Nvidia support.

4. It has a number of useful inbuilt libraries.

5. Advanced tools available for debugging.

6. Support the existing construct of C/C++

## 8. MYTH'S ABOUT CUDA

- **Myth 1**: GPUs are the only processors in a CUDA application.

  **Reality**: CPU and GPU work together in a CUDA application.

- **Myth 2**: GPUs have very wide (1000s) SIMD machines.

  **Reality**: No, a CUDA Warp is only 32 threads.

- **Myth 3**: Branching is not possible on GPUs.

  **Reality**: Incorrect, it is possible and can be implemented easily.

- **Myth 4**: GPUs are power-inefficient.

  **Reality**: Nope, performance per watt is quite good.

- **Myth 5**: CUDA is only for C or C++ programmers.

  **Reality**: CUDA can also be used with other programming languages such as JAVA, python, etc.

## 9. BENEFITS AND LIMITATIONS

### 9.1 Benefits:

1. With CUDA, high level language C can be easily used to develop applications and thus CUDA provides flexibility.

2. GPU provides facility that ample number of threads can be created concurrently using minimum number of CPU resources.

3. CUDA provides a considerable size of shared memory (16KB). This is one of the fast shared memories that can be shared among threads so that they can communicate with each other.

4. Full support for integer and bitwise operations.

5. Compiled code will run directly on GPU.

## 9.2 Limitations:

1. No support of recursive function. We have to implement recursion functions with the help of loops.

2. Many deviations from Floating Point Standard (IEEE 754) i.e.: Tesla does not fully support IEEE spec for double precision floating point operations.

3. No texture rendering.

4. CUDA may cause GPU and CPU bottleneck that is because of latency between GPU and CPU.

5. Threads should only be run in groups of 32 and up for best performance i.e.: 32 stays to be the magic number.

6. The main limitation is only supported on NVidia GPUs.

## 10. APPLICATIONS

CUDA provided benefit for many applications. Here list of some [10]:

- Seismic Database - 66x to 100x speedup

- Molecular Dynamics - 21x to 100x speedup

- MRI processing - 245x to 415x

- Atmospheric Cloud Simulation - 50x speedup.

1. **Fast video transcoding** -raw computing power of GPUs can be harnessed in order to transcode video much faster than ever before.[11]

2. **Video Enhancement-**Complicated video enhancement techniques often require an enormous amount of computations. Ex: ArcSoft was able to create a plug-in for its movie player which uses CUDA in order to perform DVD up scaling in real time![15]

3. **Computational Sciences-**In the raw field of computational sciences, CUDA is very advantageous. For example, it is now possible to use CUDA with MATLAB, which can increase computations by a great amount. Other common tasks such as computing eigenvalues, or SVD decompositions, or other matrix mathematics can use CUDA in order to speed up calculations.[15]

4. **Medical Imaging**- CUDA programming is effectively used in the medical field for image processing. Using CUDA, MRI machines can now compute images faster than ever possible before

And many more applications in different domains are present.

## 11. CONCLUSION

After the comparison between CUDA and the other paradigms of parallel computing, it is clear that the future of parallel processing is very much in the hands of the NVidia's

Architecture. So what does CUDA need in order to become the API to reckon with? In a word: portability. We know that the future of IT is in parallel computing – everybody's preparing for the change and all initiatives, both software and hardware, are taking that direction. Currently, in terms of development paradigms, we're still in prehistory – creating threads by hand and making sure to carefully plan access to shared resources is still manageable today when the number of processor cores can be counted on the fingers of one hand; but in a few years, when processors will number in the hundreds, that won't be a possibility. With CUDA, Nvidia is proposing a first step in solving this problem – but the solution is obviously reserved only for their own GPUs, and not even all of them. Only the GF8 and 9 (and their Quadro/Tesla derivatives) are currently able to run CUDA programs.

Nvidia may boast that it has sold 70 million CUDA-compatible GPUs worldwide, but that's still not enough for it to impose itself as the de facto standard. All the more so since their competitors aren't standing by idly. AMD is offering its own SDK (Stream Computing) and Intel has also announced a solution (Ct), though it's not available yet. So the war is on and there won't be room for three competitors, unless another player – say Microsoft – were to step in and pick up all the marbles with a common API, which would certainly be welcomed by developers.

So Nvidia still has a lot of challenges to meet to make CUDA stick, since while technologically it's undeniably a success, the task now is to convince developers that it's a credible platform – and that doesn't look like it'll be easy. However, judging by the many recent announcements in the news about the API, the future doesn't look unpromising.

The GPUs are gaining popularity in the scientific computing community due to their high processing capability and easy availability as we have demonstrated throughout the paper, and are becoming the preferred choice of the programmers due to the support offered for programming by models such as CUDA.

## Authors

**Jayshree Ghorpade**
**M.E. [Computer],**
**MITCOE, Pune, India.**
Assistant Professor, MAEER'S MITCOE, Pune, India. 8 years of experience in IT and Academics & Research. Area of interest is Image Processing, BioInformatics & NN.

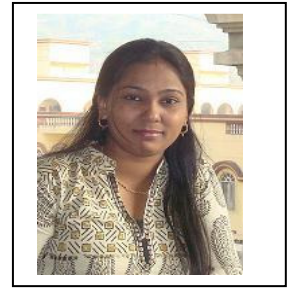

**Jitendra Parande**
B.E. [Computer]
VIT, Pune, India
Enterprise Architect
Technology Evangelist

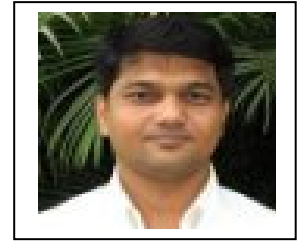

**Madhura Kulkarni**
**B.E. [Computer],**
**MITCOE, Pune, India**
 Area of interest is in software engineering, database management systems and networking

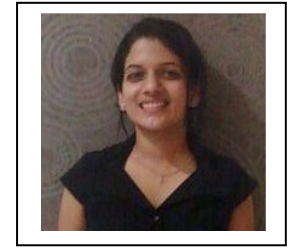

**Amit Bawaskar**
**B.E. [Computer]**
**MITCOE, Pune, India**
Area of interest is in distributed systems , networking and Artificial intelligence.

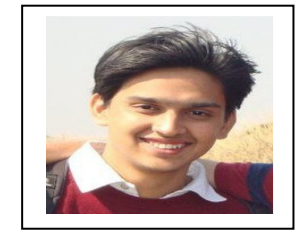